\documentclass[twocolumn,aps,pre,footinbib,floatfix,preprintnumbers,amsmath,amssymb,superscriptaddress,10pt,colorlinks]{revtex4-1}

\usepackage{ulem}
\renewcommand{\emph}{\textit}

\usepackage{selinput}
\usepackage{amsmath}
\usepackage{amsfonts}
\usepackage{amssymb}
\usepackage{graphicx}
\usepackage{dcolumn} 
\usepackage{bm} 
\usepackage{color}
\usepackage{dsfont}
\usepackage{xr}
\usepackage{mathtools}
\usepackage{here}
\usepackage[urlcolor=RoyalBlue,citecolor=RoyalBlue,linkcolor=Black,breaklinks=true]{hyperref}

\usepackage{pdfpages}
\makeatletter
\AtBeginDocument{\let\LS@rot\@undefined}
\makeatother

\renewcommand{\vec}[1]{\mathbf{ #1 }}

\definecolor{Black}{rgb}{0, 0, 0}
\definecolor{BrickRed}{rgb}{0.8, 0.25, 0.33}
\definecolor{ForestGreen}{rgb}{0.14, 0.392, 0.14}
\definecolor{RoyalBlue}{rgb}{0.149, 0.42, 0.74}
\definecolor{White}{rgb}{1, 1, 1}

\renewcommand{\paragraph}[1]{\textit{#1.}~--}
\newcommand{\balancepage}[0]{\onecolumngrid\newpage\twocolumngrid}

\begin{document}

\title{A novel approach to chemotaxis: active particles guided by internal clocks}

\author{Luis G\'omez Nava}
\thanks{L.G.N.~and R.G.~contributed equally to this work. }
\affiliation{Universit{\'e} C{\^o}te d'Azur, Laboratoire J. A. Dieudonn\'e, UMR 7351 CNRS, Parc Valrose, F-06108 Nice Cedex 02, France}

\author{Robert Gro{\ss}mann}
\thanks{L.G.N.~and R.G.~contributed equally to this work. }
\affiliation{Universit{\'e} C{\^o}te d'Azur, Laboratoire J. A. Dieudonn\'e, UMR 7351 CNRS, Parc Valrose, F-06108 Nice Cedex 02, France}
\affiliation{Institut f{\"u}r Physik und Astronomie, Universit{\"a}t Potsdam, Karl-Liebknecht-Strasse 24/25, Haus 28, 14476 Potsdam, Germany}

\author{Marius Hintsche}
\affiliation{Institut f{\"u}r Physik und Astronomie, Universit{\"a}t Potsdam, Karl-Liebknecht-Strasse 24/25, Haus 28, 14476 Potsdam, Germany}

\author{Carsten Beta}
\affiliation{Institut f{\"u}r Physik und Astronomie, Universit{\"a}t Potsdam, Karl-Liebknecht-Strasse 24/25, Haus 28, 14476 Potsdam, Germany}

\author{Fernando Peruani} 
\email{peruani@unice.fr} 
\affiliation{Universit{\'e} C{\^o}te d'Azur, Laboratoire J. A. Dieudonn\'e, UMR 7351 CNRS, Parc Valrose, F-06108 Nice Cedex 02, France}

\date{\today}

\begin{abstract}
Motivated by the observation of non-exponential run-time distributions of bacterial swimmers, we propose a minimal phenomenological model for taxis of active particles whose motion is controlled by an internal clock. 
The ticking of the clock depends on an external concentration field, e.g.~a chemical substance.
We demonstrate that these particles can detect concentration gradients and respond to them by moving up- or down-gradient depending on the clock design, albeit measurements of these fields are purely local in space and instantaneous in time. 
Altogether, our results open a new route in the study of directional navigation, by showing that the use of a clock to control motility actions represents a generic and versatile toolbox to engineer behavioral responses to external cues, such as light, chemical, or temperature gradients.
\end{abstract}

\maketitle 

\begin{figure}[!pb]
    \begin{center}
    \includegraphics[width=\columnwidth]{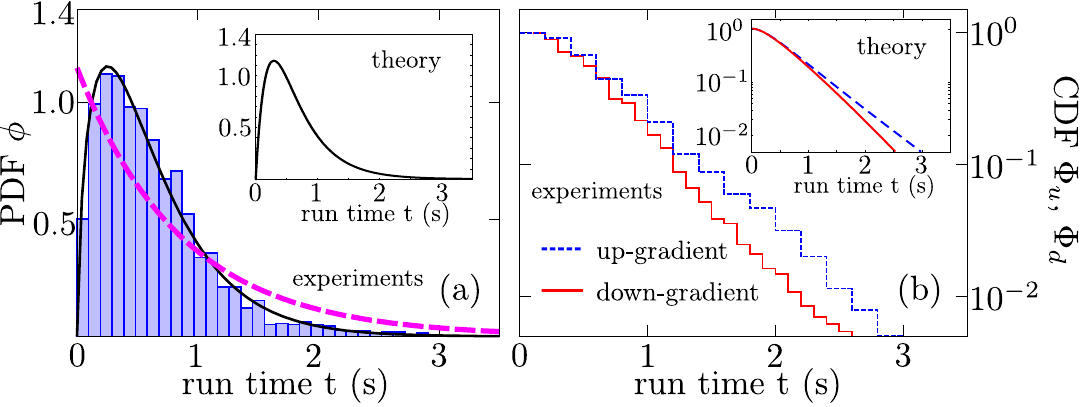}
    \end{center}
    \vspace{-0.4cm}
    \caption{Run-time distribution and chemotactic response of~\textit{P.~putida}. Left:~probability distribution function~(PDF) $\phi$ of run times. Right:~cumulative distribution function~(CDF) of run-times discriminating between up-gradient ($\Phi_u$) and down-gradient ($\Phi_d$) runs. The purple-dashed and black-solid curves in the left panel correspond to fittings with an exponential and a~$\gamma$-distribution, respectively. A~$\chi^2$-test indicates that an exponential distribution is rejected at significance level~$0.05$. The insets show that the same qualitative behavior is obtained in simulations with a clock model as introduced in the main text. Parameters: $M=2$, $\gamma_1=5 \, \mbox{s}^{-1}$, $\gamma_2(x) = [0.75 (x/L) + 0.1]^{-1} \mbox{s}^{-1}$, $L = 750 \, \mu \mbox{m}$ and $v_0 = 30 \, \mu \mbox{m}/\mbox{s}$. }
\label{fig:experiment}
\end{figure}

In the canonical picture of bacteria with run-and-tumble motility, such as {\it Escherichia coli} or {\it Salmonella}~\cite{berg1972chemotaxis,berg2008}, bacteria display exponentially distributed run-times~--~despite recent findings that suggest the possibility of noise-induced heavy-tailed distributions~\cite{Korobkova2004,Tu2005}~--~and perform chemotaxis by regulating  the associated tumbling frequency. 
By measuring the chemical concentration through clustered arrays of membrane receptors~\cite{Tindall2012} and subsequent signal processing via a complex biochemical cascade~\cite{Celani2010} that leads to an  effective memory~\cite{Schnitzer1993,Celani2010,Cates2012,Flores2012}, the bacterium is able to detect and respond to an external concentration gradient. 
It extends the duration of runs, i.e.~decreases the number of tumbles when heading in the direction of increasing attractant concentration, thus performing a biased random walk towards the attractant source.
Recently, it was discovered that several bacterial species differ from this classical picture.
Notably, in the soil bacterium~\textit{Pseudomonas putida} (\textit{P.~putida}) that displays a run-and-reverse motility pattern with multiple run modes~\cite{Hintsche2017,alirezaei2020}, the distribution of run-times is non-exponential and exhibits a refractory period, cf.~Fig.~\ref{fig:experiment}a and Refs.~\cite{Theves2013,Theves2015}; an observation that suggests that run-times are not controlled by a Poissonian mechanism~\footnote{Recent experiments with \textit{P.~putida} indicate that the nature of the run-time distribution, i.e.~an exponential vs. a non-exponential shape, is strongly influenced by the growth conditions of the bacteria:~the data presented here are obtained from bacteria, which are grown on benzoate as a carbon source, whereas in~\cite{alirezaei2020}, cells cultured in Tryptone Broth~(TB) and imaged in casamino acid gradients showed an exponential run-time distribution.}. %
By adapting the reversal statistics in response to a chemical gradient, \textit{P.~putida} is able to perform chemotaxis as indicated in Fig.~\ref{fig:experiment}b. 
However, the non-exponential run-time distribution suggests that chemotaxis in~\textit{P.~putida}, and potentially also in other bacterial species, may involve a taxis mechanism that is fundamentally different from the one reported for {\it E. coli}. 
Similar non-exponential bell-shaped run-time distributions were reported for \textit{Myxococcus xanthus}~\cite{Wu_periodic_2009} and \textit{Paenibacillus dendritiformis}~\cite{beer_periodic_2013}, which display run-and-reverse motility similar to {\it P. putida}.
Non-exponential run-times were observed also for the marine bacterium \textit{V.~alginolyticus}~\cite{xie_bacterial_2011} and, notably, also for the rotation time of the flagellar motors of \textit{Escherichia coli}~\cite{korobkova_hidden_2006}. 
Furthermore, it has been reported that surface exploration of~\textit{Escherichia coli}~\cite{ipina2019bacteria} is 
not consistent with the canonical run-and-tumble picture of bacterial motility, but involves multiple motility modes.

Motivated by the non-Poissonian run-time statistics of~\textit{P.~putida}~--~though not pretending to be a realistic, biochemical description of chemotaxis in~\textit{P.~putida}~--~we propose a minimal, phenomenological, generic model for taxis of active particles that inherently produces non-exponential run-time distributions.
Particles move at constant speed and experience velocity reversals.
The key element of the model is that the occurrence of reversal events is controlled by an internal clock.
The clock mimics the fact that the directional response of a microorganism to an external signal or field, such as a chemical concentration or temperature gradient, requires to sense this signal, internalize and process it, presumably involving cascades of biochemical events~\cite{korobkova_hidden_2006,wang_noneq_2017}, and to execute a behavioral response~(e.g.~a reversal), after which the microorganism continues sensing, processing and responding to the signal. 
We assume that this complex cycle can be reduced to a series of stochastic checkpoints or steps, where some or all of them depend on the external signal.
These steps are represented by the ``ticks''~of a clock. 
The transitions between two consecutive ticks are modeled as Poissonian processes with concentration-dependent transition rates. 
Ergo, the clock dynamics is, by definition, Markovian:~it does not incorporate or presuppose memory in any manner. 
On the other hand, the distribution of the times between two consecutive behavioral responses is non-exponential as observed in~\textit{P.~putida} experiments, cf.~Fig.~\ref{fig:experiment}a. 
Experimental details on cell culturing, the chemotaxis chamber and imaging can be found in the Supplemental Material~(SM) and Refs.~\cite{Sambrook2001,Harwood1990,harwood_aromatic_1984,Pohl2017,Theves2013}. 
In this study, we demonstrate that the design of the clock controls the long-time motility of the particles: some of the clock designs result in signal-insensitive particles and a variety of other designs lead to particles displaying actual taxis. 
The taxis responses include either up-gradient or down-gradient biased motion. 
In short, we show how a clock can be used to guide active particles subjected to external stimuli, and to obtain non-exponential run-time distributions.

%

\begin{figure}[!pt]
    \begin{center}
    \includegraphics[width=\columnwidth]{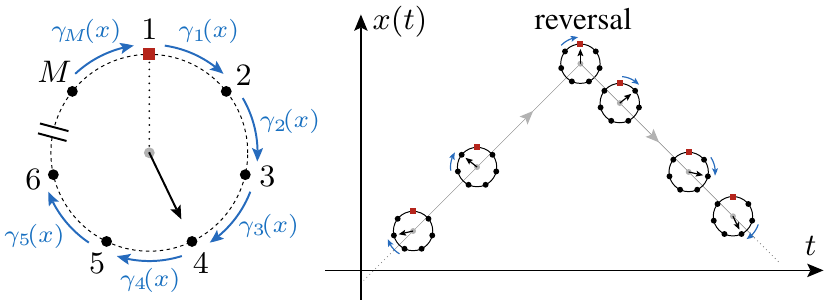}
    \end{center}
    \vspace{-0.5cm}
    \caption{Illustration of the internal clock dynamics~(left panel) and the resulting particle motion~(right panel). The transition rates~$\gamma_{n} \! \left( x \right) = f_n[c(x)]$ may be concentration dependent. A particle reverses its direction of motion, i.e.,~$s(t) \rightarrow -s(t)$, after a full clock cycle has been completed as indicated by a red square on the clock. A segment of a space-time trajectory is represented by a gray line in the right panel. Additionally, the internal dynamics is overlaid at those instances in time in which clock ticks occur. }
\label{fig:scheme}
\end{figure}

\paragraph{Model} 
We consider active particles that move at constant speed~$v_0$ in a one-dimensional system of size~$L$ with reflecting boundary conditions. 
Particles are exposed to a temporally constant external field~$c(x)$. The equation of motion of one of these particles is given by   
\begin{equation}
	\frac{d x(t)}{dt} = v(t) =  v_0 s(t),
\label{eqn:equation_of_motion}
\end{equation}
where~$x(t)$ denotes the position of the particle and~$s(t)\in \{ -1 , 1 \}$ indicates its direction of motion. 
The variable $s(t)$ undergoes stochastic transitions such that~$s(t) \! \to \! -s(t)$. 
These reversal events are controlled by a stochastic $M$-tick clock.  
A reversal occurs every time the clock completes a full cycle, i.e.~in the transition from $M$ to $1$, as illustrated in Fig.~\ref{fig:scheme}. 
Thus, the state of a particle is characterized by its position~$x(t)$, its orientation~$s(t)$ as well as the internal state~(or tick) of the clock~$n \!\in \! \{1,2,\dots,M\}$.

\paragraph{Clock dynamics} 
The transition between tick $n$ to $n+1$ is modeled by a Poisson process with a concentration dependent transition rate~$\gamma_{n} \! \left( x \right)=f_n[c(x)]$, where the function~$f_n[c]$ denotes the dependence of the $n$-th rate on the external field~$c(x)$. 
For simplicity, we will henceforth denote the rates as functions of~$x$, keeping in mind, however, that the dependence arises via the gradient in~$c(x)$.  
If all transition rates are independent of the external field~$c(x)$ and equal, i.e., $\gamma_n=\beta$ with a positive constant~$\beta$, the spatially homogeneous case studied in~\cite{grossmann_diffusion_2016} is recovered. 
It is important to notice that the model can be easily formulated in two dimensions as explained in the SM, however, we focus on the one-dimensional scenario for simplicity and without loss of generality here. 
Let $P_{n}(t)$ denote the probability to find the clock in state~$n$ at time~$t$.
The temporal evolution of $P_{n}(t)$ can be expressed via the Master equation~\cite{gardiner_stochastic_2010}
\begin{subequations}
\begin{align}
    \frac{dP_1(t)}{dt} &= -\gamma_1\!\!\:\big( x(t) \big) P_1 \! \left( t \right) + \gamma_{M} \!\!\: \big( x(t) \big) P_{M}(t) \\
    \frac{dP_n(t)}{dt} &= -\gamma_n \!\!\: \big( x(t) \big) P_n(t) + \gamma_{n-1} \!\!\: \big( x(t) \big) P_{n-1}(t)
\end{align}
\end{subequations}
with~$n \! \in \! \left\{ 2,3,\dots,M \right\}$. 
Notice that this is a closed chain of states.

\paragraph{Run-time distribution}
To compute the run-time distribution, we solve for the first passage time of a directed walk from state~$1$ to state~$M$ within the clock. 
Let us consider a particle that moves in direction $s_0$ at time $t_0$ and is in state $1$. We follow its dynamics in space given by $x \! \left( t \right) \!=\! x_0 \!+\! s_0 v_0 (t-t_0)$ and stop it when it leaves state~$M$. 
We can obtain the probabilities~$P_{n}(t)$ by recursion:
\begin{subequations}
\label{eqn:run-timepdf}
    \begin{align}
    \!\! P_1 \! \left( t \right) &\!=\! e^{- \! \int_{t_{0}}^{t} dt' \gamma_1 \!\!\: \left ( x(t') \right )} \!\!\: , \!\! \\
    \!\! P_n \! \left( t \right) &\!=\!\! \int_{t_0}^{t} \! dt'' \gamma_{n-1} \!\!\: \big ( x(t'') \big ) P_{n-1} \! \left( t'' \right) e^{- \! \int_{t''}^{t} dt' \gamma_n \!\!\: \left ( x(t') \right )} \!\!\:. \! \!
    \end{align}	
\end{subequations}
The run-time distribution is given by~$\phi(t| x_0, s_0) \!=\! \gamma_M \big( x(t) \big) P_M(t)$; an example of $\phi$ for a $M=2$ clock is shown in Fig.~\ref{fig:experiment}.

\paragraph{Spatio-temporal dynamics}
Since we are dealing with a genuine Markov process, the exact, full spatio-temporal dynamics of the problem can be described in terms of a Master equation for the probability density~$P_n^{\pm} \! \left( x,t \right)$ to find a particle in position~$x$ at time~$t$ oriented along the direction~$s \!=\! \pm 1$ with the internal state~$n$:
\begin{align}
	\Big [ \partial_t \!\pm \! v_0 \partial_x \!\!\:\!+\! \gamma_{n} \! \left( x \right) \! \Big ] \! P_{n}^{\pm} \! \left( x,t \right)  &\!=\!
	\begin{cases}
	  	\!\:\! \gamma_{M} \! \left( x \right) \!  P_{M}^{\mp} \! \left( x,t \right) \! , & \!\!\!\! n = 1, \\
	  	\!\:\! \gamma_{n-1} \! \left( x \right) \!  P_{n-1}^{\pm} \! \left( x,t \right) \! , & \!\!\!\! n \ge 2. 
	\end{cases}
\label{eqn:master_components} 
\end{align}
This is a system of~$2M$ coupled linear partial differential equations. 
Notice that the dynamics of $P_1^{\pm} \! \left( x,t \right)$ is special since it contains the dynamics of reversals. 
Eq.~\eqref{eqn:master_components} can be expressed in a concise matrix form for the vector
\begin{align*}
	\vec{P}\!\left( x,t \right) = \left( P_1^{+} \! \left( x,t \right)\!,P_1^{-} \! \left( x,t \right)\!,\dots,P_M^{+} \! \left( x,t \right)\!,P_M^{-} \! \left( x,t \right) \right)^{\!T}
\end{align*}
as follows: 
\begin{align}
	\partial_t \vec{P} \! \left( x,t \right) = \mathcal{C} \cdot  \partial_x \vec{P} \! \left( x,t \right) + \mathcal{L}(x) \cdot \vec{P} \! \left( x,t \right) \!, 
 \label{eqn:master_matrix}
\end{align}
where the matrix~$\mathcal{C}$, which encodes the motility of particles, is diagonal with the coefficients~$\mathcal{C}_{ij} \!=\! (-1)^{j} v_0 \delta_{ij}$ 
and~$\mathcal{L}\!\left( x \right)$ contains the transition rates of the clock dynamics as well as the reversals. 
The time-dependent solution may be written as~$\vec{P} \! \left( x,t \right) \!=\! \exp \left[ \left( t-t_0 \right) \!\mathcal{M} \! \left( x \right) \right] \! \vec{P} \! \left( x,t_0 \right)$, where we introduced the operator~$\mathcal{M} \! \left( x \right)  \! = \! \mathcal{C} \partial_x \!+\! \mathcal{L} \! \left( x \right)$ on the right hand side and~$\vec{P} \! \left( x, t_0\right)$ abbreviates the initial condition~\cite{risken_fokker-planck_1996}.

\paragraph{Clock design \& taxis response}
If there is an observable taxis response, the density $\rho(x,t) = \sum_{n=1}^{M} \left[P_n^{+}(x, t)+P_n^{-}(x, t)\right]$ to find a particle at position~$x$ should be spatially modulated in the long time limit, i.e., $\rho(x)=\lim_{t\to\infty} \rho(x,t)$ should not be constant. 
We exploit the fact that the time-independent solution of the Master equation~\eqref{eqn:master_components} is unique~\cite{kampen_stochastic_2011}.  
In this way, we can easily verify whether a non-constant~$\rho(x)$ should be expected for a given clock design. 
We classify clock designs into two categories, homogeneous and inhomogeneous clocks, addressed separately in the following.

{\it (1) Homogeneous clocks:} 
We call a clock~\textit{homogeneous} if all transition rates~$\gamma_n(x)$ are identical functions of the concentration:~$\gamma_n \! \left( x \right) \!=\! f[c \! \left( x \right)]$ for~$n=1,2,\dots,M$ with an arbitrary function~$f[c]$.  
We start by considering a special case of great relevance: a clock model with only one tick~($M=1$). 
In this particular case, the Master equation reduces to 
\begin{align}
	\Big ( \!\!\:\partial_t \pm v_0 \partial_x \!\!\:\Big ) \!\:\! P_1^{\pm} \! \left( x,t \right) = \gamma_1 (x) \Big[ P_1^{\mp} \! \left( x,t \right) - P_1^{\pm} \! \left( x,t \right) \!\!\: \Big] .
\label{eqn:one_tick_dyn}
\end{align}
Even though the transition rate~$\gamma_1$ depends on $c(x)$, the stationary solution is spatially independent:~$P_1^{\pm} \! \left( x \right) \!=\! (2L)^{-1}$.
In short, a particle moving at constant speed is unable to detect a chemical gradient if the reversal process is controlled by a Poisson process ($M=1$) with a concentration dependent transition rate. 
Surprisingly, a homogeneous stationary solution,~$P_n^{\pm} \! \left( x \right) \!=\! \left( 2ML \right)^{-1}$, is also obtained for all homogeneous clocks with $M>1$, despite the run-time distributions are $\gamma$-shaped.
\begin{figure*}[ht!]
    \begin{center}
    \includegraphics[width=0.95\textwidth]{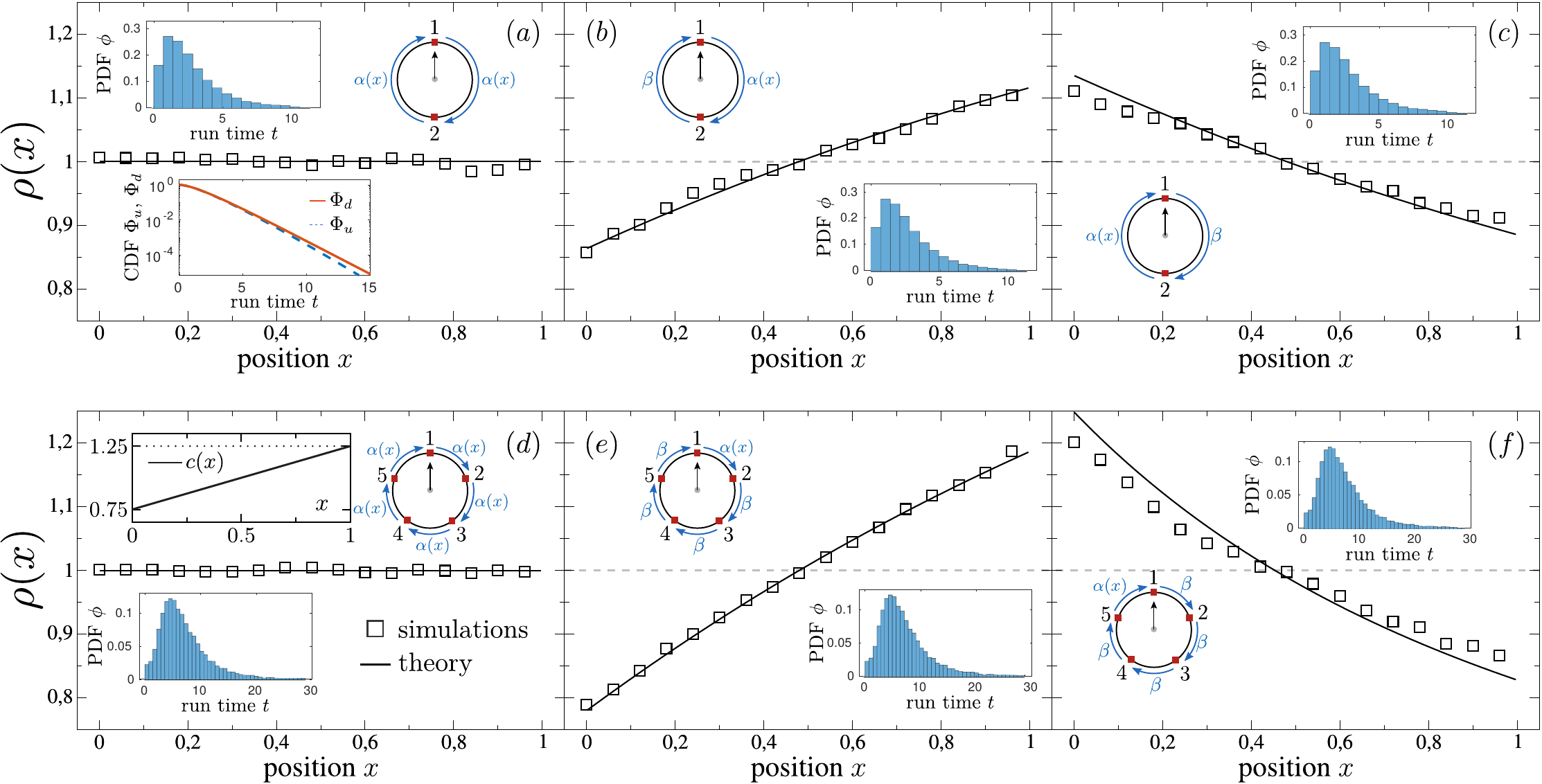}
    \end{center}
    \vspace{-0.5cm}
    \caption{Stationary distributions $\rho(x)$ for two homogeneous, $(a)$ and $(d)$, and four inhomogeneous clocks, $(b)$,$(c)$, $(e)$ and $(f)$, with $M=2$ and $M=5$ internal states. Points represent particle-based simulations and lines show the density profiles as predicted by the perturbation theory. The external field $c(x)$, common for all cases shown, is displayed as an inset in panel~(d). If all rates are equal~(homogeneous clocks), a uniform density profile develops~--~particles are thus unable to respond to the chemical gradient~[panels~$(a)$ and $(d)$]. However, there can be a run-time bias, calculated from Eq.~\eqref{eqn:run-timepdf} as indicated in the inset of panel~$(a)$. A nonuniform density profile is observed if the symmetry of the internal dynamics is broken. The probability to find a particle is proportional to the modulation of the first transition rate~$\alpha(x)$~[panels~$(b)$ and $(e)$]. By inverting the design of the clock~[panels~$(c)$ and $(f)$], the gradient in the resulting density profile switches sign with respect to the former case. For all cases, the run-time distribution $\phi(t)$ (shown in each panel as an inset using the same bin width of $\Delta t = 0.7$) is bell-shaped when there is more than $M = 1$ internal state. Note that the shape of the run-time distribution as well as the steepness of the density gradient depend on the design of the clock. Parameters in arbitrary units:~$\alpha(x) \!=\! c(x)$, $c(x) = 0.75 + 0.5x$, $\beta = 1$, $v_0 = 0.01$, $L = 1$. For particle-based simulations, $N = 10^4$ particles were simultaneously tracked using a stochastic Euler scheme~\cite{gardiner_stochastic_2010} with~$\Delta t = 10^{-3}$ and density distributions were averaged over time neglecting initial transients. }
\label{fig:profiles}
\end{figure*}
We stress that homogeneous clocks will, however, lead to a measurable run-time bias at the single particle level, i.e.~run-times up-gradient differ compared to their down-gradient counterparts, cf.~the inset in Fig.~\ref{fig:profiles}a and Eq.~\eqref{eqn:run-timepdf}. 
Nevertheless, there is no accumulation at concentration maxima nor minima. 
In short, homogeneous clocks with multiple ticks do not induce a long-time taxis response as illustrated in Fig.~\ref{fig:profiles}a,d. 
We thus conclude that the measurement of stationary concentration profiles is an indispensable piece of information, whereas the observation of a run-time bias in the individual trajectories provides only insufficient insight into the long-time chemotaxis performance.

{\it (2) Inhomogeneous clocks:}
If at least two transition rates are different from one another, the stationary state is not spatially homogeneous and nontrivial stationary profiles of $\rho(x)$ can develop~(see Fig.~\ref{fig:profiles}b,c,e, and~f). 
For arbitrary spatial dependencies of the transition rates~$\gamma_n(x)$, we cannot find the exact stationary solution of Eq.~\eqref{eqn:master_components} analytically, but various approximation methods can be applied. 
In Ref.~\cite{nava_markovian_2018}, a drift-diffusion approximation of the Fokker-Planck type~\cite{gardiner_stochastic_2010}
\begin{align}
\label{eqn:structFP1}
	\partial_t \rho \! \left( x,t \right) \simeq - \partial_x \Big[ f(x) \rho \! \left( x,t \right) \Big ] \!  + \partial_x^2 \Big [ D(x) \rho \! \left( x,t \right) \Big ] 
\end{align}
for the long-time dynamics of the density~$\rho(x,t)$ was proposed in a related context, derived under the assumption that the mean distance traversed by a particle in between two reversals is shorter than the characteristic length scales at which the chemical gradient varies.
The approximation scheme is based on a slow mode reduction of the full Master equation~[Eqs.~\eqref{eqn:master_components}]:~as the particle density~$\rho(x,t)$ is a conserved quantity, its dynamics is slow, thus enabling the systematic adiabatic elimination of fast degrees of freedom. 
In this way, drift~$f(x)$ and diffusion~$D(x)$ can be analytically obtained for any clock motif~\cite{nava_markovian_2018}. 
This approximation allows the prediction of the density profile~$\rho(x)$ in the long-time limit based on the stationary solution of Eq.~\eqref{eqn:structFP1}. 

\begin{figure}[tb]
    \begin{center}
    \includegraphics[width=\columnwidth]{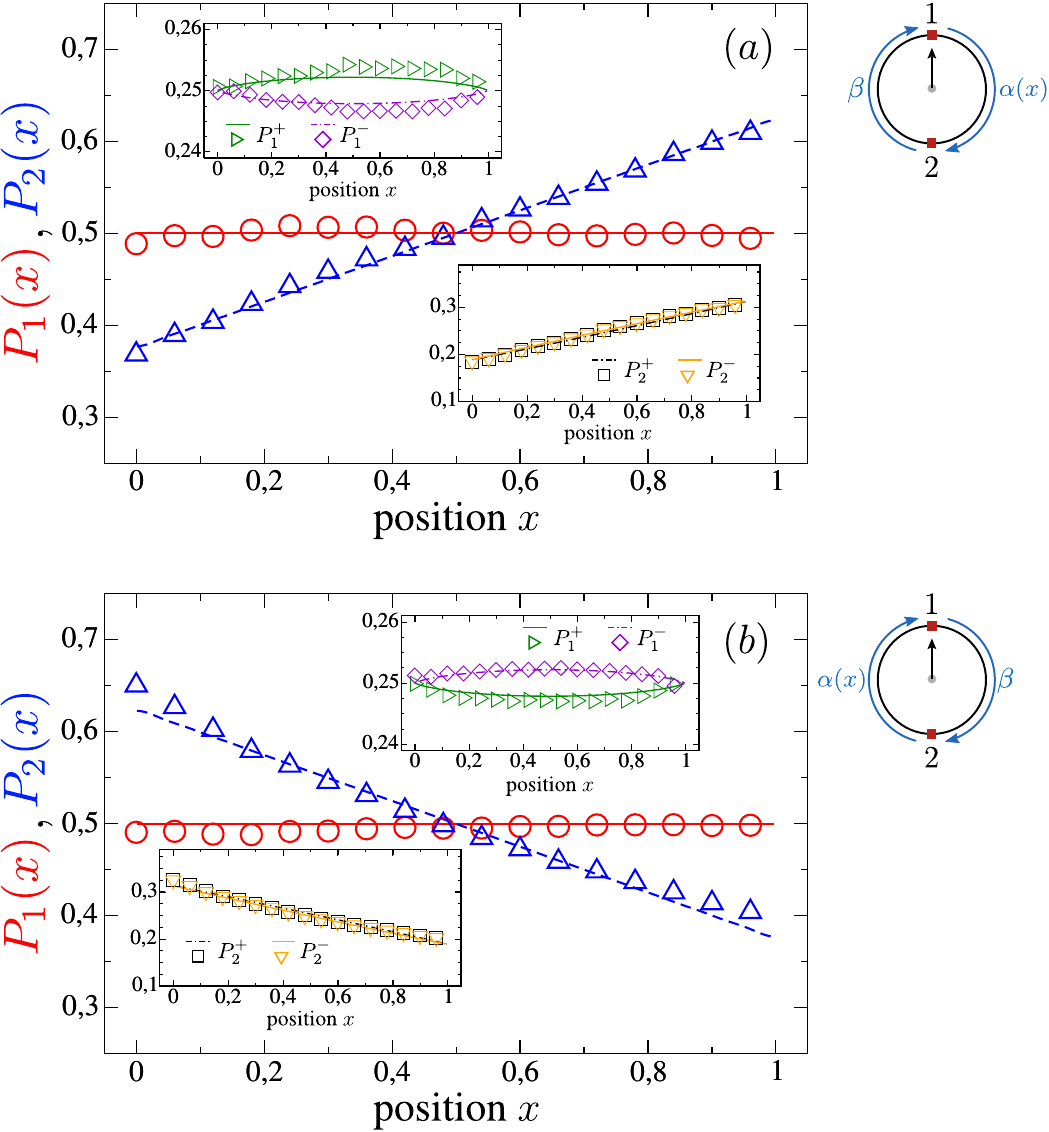}        
    \end{center}
    \vspace{-0.45cm}
    \caption{Comparison of the prediction of the perturbation theory and individual-based simulations at the level of the probabilities~$P_i(x)=P_i^{+}(x)+P_i^{-}(x)$ to find a particle in state~$i$ at position~$x$ in the stationary state for the clock motifs shown in Fig.~\ref{fig:profiles}b,c. The insets indicate the splitting of these probabilities into up- and down-gradient motion, i.e.~the probability densities~$P_i^{\pm}$. Lines correspond to predictions of the perturbation theory, points show individual-based simulations. Parameters are identical to Fig.~\ref{fig:profiles}. }
\label{fig:components}
\end{figure}

In this work, we present a perturbative approach that enables us to predict the stationary solution for the vector~$\vec{P}(x)$ including all of its components beyond the scalar, stationary density~$\rho(x)$.  
In Figs.~\ref{fig:profiles} and \ref{fig:components}, we compare particle-based simulations with the perturbative solution whose derivation is sketched below~(see~SM for additional technical details). 
The central idea of the perturbation theory is that the transition rates are only weakly modulated in space allowing us to split the matrix~$\mathcal{L} \! \left( x \right)$ into a constant and a spatially dependent part, $\mathcal{L} \! \left( x \right) \!\!=\! \mathcal{L}_{0} \!+ \! \varepsilon \mathcal{L}_{\Delta} \! \left( x \right)$, where the second matrix is defined in such a way that~$\int_{0}^{L} dx \, \mathcal{L}_{\Delta} \! \left( x \right) \!=\! 0$. Now, we assume that~$\varepsilon$ is a small parameter allowing to expand the stationary solution as a power series:~$\vec{P} \! \left( x \right) \!=\! \sum_{n=0}^\infty \vec{P}_{\mu} \! \left( x \right) \varepsilon^{\mu}$. Inserting this ansatz into Eq.~\eqref{eqn:master_matrix} and collecting orders in~$\varepsilon$ yields a systematic way to construct the stationary solution. 
This approach allows us to demonstrate (\textit{i})~that active particles controlled by inhomogeneous clocks display a taxis response and  (\textit{ii})~how the clock design determines the type of response. 
Figures~\ref{fig:profiles}b and \ref{fig:profiles}c show the emerging density profiles for particles with $M=2$-clocks subjected to the same external field, where the transition rates are $\gamma_1 = \alpha(x)$ and $\gamma_2 = \beta$ in Fig.~\ref{fig:profiles}b, while $\gamma_1 = \beta$ and $\gamma_2 = \alpha(x)$ in Fig.~\ref{fig:profiles}c. 
The important observation here is that~$\rho(x)$ increases monotonically for the former clock and decreases for the latter one as one moves in up-gradient direction with respect to~$c(x)$; notably, the average tumbling frequency increases in both cases in the up-gradient direction. 
In Fig.~\ref{fig:profiles}e and \ref{fig:profiles}f, a similar scenario is shown for an $M=5$-clock for comparison.
While the qualitative trends remain the same, the gradient in the nonuniform density profiles becomes steeper in the $M=5$-case.
In Fig.~\ref{fig:components}, a detailed comparison of individual-based simulations and the perturbation theory is shown for two clocks with two ticks~($M=2$), confirming that the presented perturbative approach can indeed predict not only the overall density profile~$\rho (x)$ but also the individual probability densities~$P_i^{\pm}(x)$ to find a particle in state~$i$ at position~$x$ moving up- or down-gradient, respectively, in the stationary state.

\begin{figure}[!b]
    \begin{center}
    \includegraphics[width=\columnwidth]{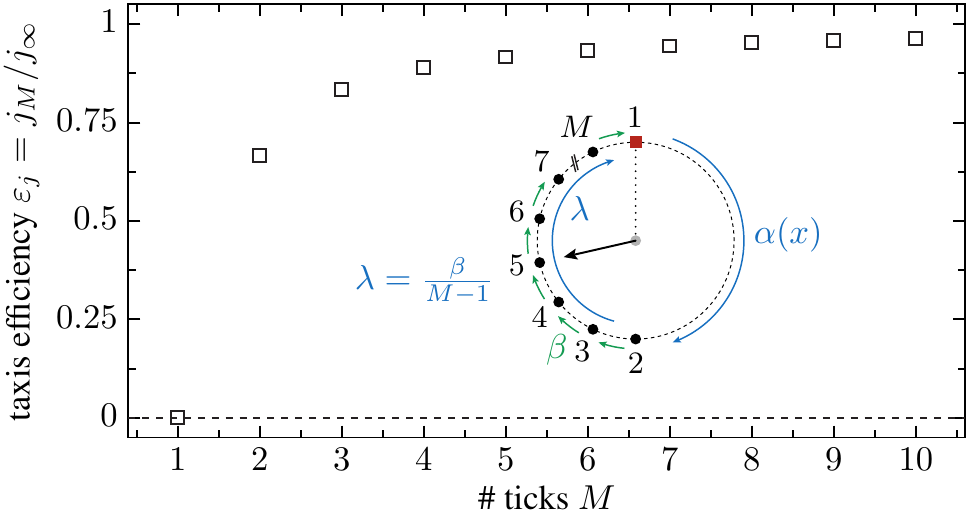}
    \end{center}
    \vspace{-0.5cm} 
    \caption{Taxis efficiency, characterized by the relative current~$\varepsilon_{j} = j_M / j_{\infty}$, see text and Eq.~\eqref{eqn:rel_curr}, as a function of the number of ticks $M$, which measures the accuracy of the clock. The mean frequency~$\lambda$ for the occurrence of the last~$M\!-\!1$ ticks was fixed to be independent of~$M$ by choosing~$\beta = (M-1) \lambda$. The efficiency of sensing an external field increases with the accuracy of the clock. There is no chemotactic response for~$M=1$. Parameters in arbitrary units~(cf.~Fig.~\ref{fig:profiles}): position of the current measurement $\bar{x} = 0.5$, $\alpha(x)$ is a linear function with $\alpha(x) = 0.75 + 0.5 x$, $\lambda \!=\!1$, $v_0 \!=\! 0.01$, $L = 1$. }
\label{fig:efficiency}
\end{figure}

\paragraph{Efficiency of taxis response}
To quantify the efficiency of the taxis response of particles whose reorientation is controlled by clocks, we measure the emerging particle current that arises once a chemical gradient is instantaneously switched on at~$t=0$, given an initially flat density distribution in a homogeneous environment for~$t\le0$. 
We focus on the role of the number of ticks~$M$ for clocks where $\gamma_1(x)=\alpha(x)$, while all~$\gamma_n = \beta$ with~$n>1$ are constant and identical. 
In order to make cases with different numbers of ticks comparable, we fix the mean time~$\lambda^{-1}$ spent by the clock between ticks~$2$ and~$M$ by setting~$\gamma_n = \beta = \lambda (M-1)$, where $\lambda$ is a constant. 
In this way, clocks with larger~$M$ correspond to more accurate clocks in the sense that the standard deviation of the time between ticks~$2$ and~$M$ divided by its mean value scales as~$1/\sqrt{M-1}$.
Applying Eq.~\eqref{eqn:structFP1} to an initially homogeneous density distribution~($\rho(x,t=0) = \rho_0$), we derive the initial current
\begin{subequations}
\label{eqn:rel_curr}
    \begin{align}
	    \hspace*{-0.1cm} j_M(x,t=0)  &= \rho_0 \Big [ f(x) - \partial_x D(x)  \Big ] , \\ 
    & \hspace*{-1.6cm} = \rho_0 (M-1) \!\!\: \cdot \!\!\: \frac{v_0^2}{2} \!\!\: \cdot \!\!\: \frac{2 \beta + (M-2) \alpha(x) }{\alpha(x) \big [  \beta + (M-1) \alpha \big ]^2 } \!\!\: \cdot \!\!\: \frac{d \alpha(x)}{dx},
    \end{align}
\end{subequations}
where the first line is the general expression and the second line follows for the particular clock model under consideration, derived by application of the drift-diffusion approximation proposed in~\cite{nava_markovian_2018}. 
We obtain~$j_1=0$ for clocks with only one tick. 
Fig.~\ref{fig:efficiency} shows that the ratio~$\varepsilon_{j} = j_M / j_{\infty}$ increases above zero only for~$M>1$~--~an observation that confirms that at least two ticks are required to obtain a non-trivial response to the external field. 
Furthermore, we find that the taxis efficiency increases with the clock accuracy, saturating for large values of $M$.  
Note that the asymmetry of initial density currents is compensated by a nonuniform density distribution of particles in the steady state.

\paragraph{Concluding remarks}
We studied the ability of active particles to perform taxis controlled by internal clocks. 
Our results reveal that clock designs with homogeneous transition rates~$\gamma_n(x)$, i.e., $\gamma_n(x) = f[c(x)]$ for all $n$ with an arbitrary function~$f[c]$, cannot generate a taxis response.   
It is important to realize that~--~despite the absence of a taxis response~--~the frequency of reversals may depend on the particle position, increasing or decreasing up-gradient depending on the clock design.
We have shown that the clock has to fulfill the following requirements in order to generate a taxis response:~(\textit{i}) a number of ticks $M\geq2$, (\textit{ii}) at least one transition rate must depend on the external field, i.e., depends on~$x$, and (\textit{iii}) inhomogeneous transition rates, i.e., some or all transition rates should be pairwise distinct.
Regarding the efficiency of the taxis response, we found that it increases with~$M$, i.e.~with the clock accuracy, though saturating for large values of~$M$. 
We stress that the obtained taxis responses are not due to the fact that we focused on velocity-reversing particles.  
All results hold true if we replace velocity reversals by tumbling events:~if~--~instead of reversing with probability one~--~the particle reverses its direction of motion with probability~$p$ and keeps it otherwise, all results reported here hold qualitatively true~(cf.~SM). 
By putting velocity reversals and tumbling events on an equal footing, it becomes evident that the crucial difference between the chemotactic mechanism of run-and-tumble bacteria, such as {\it E. coli}, and the one proposed here is the fundamentally distinct dynamics of the motility-control mechanisms, and not the type of turning maneuver~(a tumble or a reversal) which is activated by the control mechanism. 
The two most compelling differences between these two mechanisms can be summarized as follows:
(a) In a uniform external field, i.e.~in the absence of a gradient, run-and-tumble bacteria display an exponential distribution of run-times, and thus the temporal sequence of tumbling events is well-characterized by a single rate. 
In the here-proposed model, which is phenomenologically consistent with observations of run-and-reverse bacteria such as~\textit{P. putida}, the distribution of run-times is never exponential for~$M>1$, but $\gamma$-shaped, cf.~Fig.~\ref{fig:experiment}. 
This implies that  the temporal sequence of reversals (or tumble events)  cannot be parametrized by just a single rate, an observation indicating that the dynamics that controls the triggering of such events is more complex than a simple Poisson process~\cite{korobkova_hidden_2006,wang_noneq_2017}; we show that it may well be represented by a clock model. 

(b) The canonical chemotaxis strategy of run-and-tumble bacteria relies on a memory: when experiencing a temporal increase in chemoattractant concentration, they decrease their tumble frequency. 
In the proposed clock-controlled taxis mechanism, in contrast to the classical chemotaxis picture, biased up-gradient motion can occur even when the tumbling frequency increases as the bacterium moves up-gradient.  
Moreover, depending on the clock design, we have observed (\textit{i}) absence of taxis, (\textit{ii}) up-gradient motion or (\textit{iii}) down-gradient motion, even though the tumbling frequency was increasing with increasing concentration in all three cases, as illustrated in Fig.~\ref{fig:profiles}. 
This highlights the relevance of the clock architecture and indicates that the taxis direction is not dictated by the concentration dependency of the tumbling frequency, but rather by the clock design.   
In other words, we conclude from our results that it is generally not sufficient to measure the run-time bias only in order to infer the chemotaxis strategy of a microogranism but the measurement of the stationary concentration profile in a chemical gradient provides necessary, complementary information.
We recall in this context that homogeneous clocks, where all transition rates are equal, imply a run-time bias at the single-particle level but will not yield a nonuniform concentration profile, i.e.~particles will not accumulate at concentration maxima nor minima. 
Note also that the measurement of these profiles is experimentally managable, e.g.~via microfluidic maze structures as reported recently~\cite{salek_bacterial_2019}. 
Altogether, the proposed clock-controlled taxis mechanism is a powerful conceptual toolbox that allows us to engineer a large variety of taxis responses, which may also play a key role in the design of simple robots~\cite{mite2016} or for controlled assembly of active colloids~\cite{bauerle_self_2018,karani_tuning_2019}.
While the proposed phenomenological model is qualitatively consistent with experimental observations of~\textit{P.~putida}, it is important to stress that we currently have no further mechanistic evidence that~\textit{P.~putida} or other bacteria operate by such a clock. 
We insist on the phenomenological nature of the proposed clock model for sensing a signal, its internalization and processing, presumably involving cascades of biochemical events depicted by stochastic checkpoints. 
Identifying the intracellular mechanisms regulating the chemotactic responses of~\textit{P.~putida} as well as other bacteria is a long-term experimental challenge beyond the scope of our current work. 
We thus hope that this study will open the door to a new series of experimental and theoretical works that advance our understanding of the directional navigation of bacteria.

\begin{acknowledgements}
L.G.N., R.G.~and F.P.~acknowledge financial support from Agence Nationale de la Recherche via Grant No. ANR-15-CE30-0002-01. 
L.G.N. was additionally supported by CONACYT PhD scholarship 383881 and    
R.G. by the People Programme (Marie Curie Actions) of the European Union's Seventh Framework Programme 
(FP7/2007-2013) under REA grant agreement n. PCOFUND-GA-2013-609102, through the PRESTIGE programme coodinated by Campus France. 
M.H.~and C.B.~thank the research training group GRK 1558 funded by Deutsche Forschungsgemeinschaft for financial support. 
\end{acknowledgements}


%

\balancepage


\cleardoublepage



\section*{Supplementary material}

# SUPPLEMENTAL MATERIAL
# A novel approach to chemotaxis: active particles guided by internal clocks

Luis Gómez Nava, Robert Großmann, Marius Hintsche, Carsten Beta, and Fernando Peruani

### Appendix A: Experiments with P. putida

#### 1. Cell culture

*P. putida* KT2440 was streaked on 1.5% agar (AppliChem, Germany) containing LB medium (AppliChem, Germany) and grown at 30 °C. A single-colony isolate was used to inoculate 10 ml of M9 medium [1] supplemented with 5 mM sodium benzoate as a carbon source [2, 3]. The culture was grown for 36 h on a rotary shaker (300 min$^{-1}$, 30 °C) to stationary phase. It then was diluted 1:100 into 25 ml of fresh M9 medium and grown 12 h to an OD$_{600}$ of approximately 0.3 in exponential phase. Bacteria were washed three times by centrifugation at 1000g for 10 min and carefully resuspended in 10 ml motility buffer ($1 \times 10^{-2}$ M potassium phosphate, $6.7 \times 10^{-2}$ M NaCl, $1 \times 10^{-4}$ M EDTA and 0.5%w/v glucose; pH 7.0). Cells were diluted further to an OD$_{600}$ of 0.05 before filling them into the chemotaxis device.

#### 2. Microfluidics and imaging

We used a $\mu$-Slide Chemotaxis 3D (ibid., Martinsried, Germany) to generate stable linear gradients of benzoate. Filling was performed as detailed in [4] with concentrations in the source and sink reservoir of 0.5 mM and 5 mM sodium benzoate, respectively. Imaging was done using an IX71 inverted microscope with a 20× UPLFLN-PH objective (both Olympus, Germany) in phase contrast mode with an attached Orca Flash 4.0 CMOS camera (Hamamatsu Photonics, Japan). Video data was acquired for two minutes at 20 Hz, approximately one hour after filling the channel. The gradient linearity and stationarity was checked previously [4]. The viewport and focal plane were set in the center of the channels observation region, about 35 µm from top and bottom. Image processing and tracking was performed as previously described in [4].

#### 3. Filtering and smoothing

To filter out damaged, non-motile cells and those swimming on very wobbly trajectories, we discarded tracks based on a number of criteria. Tracks below a mean speed of 5 µm s$^{-1}$, longer than 20 s, above the 80th percentile of median curvature and with a total displacement below 5 µm are discarded. To further smooth these tracks, we applied a 3 point, second order Savitzky-Golay filter. Because tracks often begin or end with a reversal we cut of the first and last 0.5 s of each recorded track.

#### 4. Run-and-tumble recognition

Characterization of reversal events was performed with the algorithm of Theves *et al.* [5]. This algorithm determines reversals by evaluating local extrema in the speed and turn rate time series and counting a reversal for sufficiently deep peaks and troughs.

### Appendix B: Active particles with clock-controlled tumbling in one and two dimensions

As indicated in the main text, the appearance of a stationary profile $\rho(x)$ does not depend on the response triggered by the clock, i.e., whether this is a velocity reversal or a tumbling event. Here, we show first – using individual based simulations – that the same qualitative results, described in the main text, are recovered if reversals are replaced by tumbling events. We also explain how the model can be generalized to two dimensions and provide numerical evidence that indicates that the clock mechanism is also effective to produce a taxis response in two dimensions.

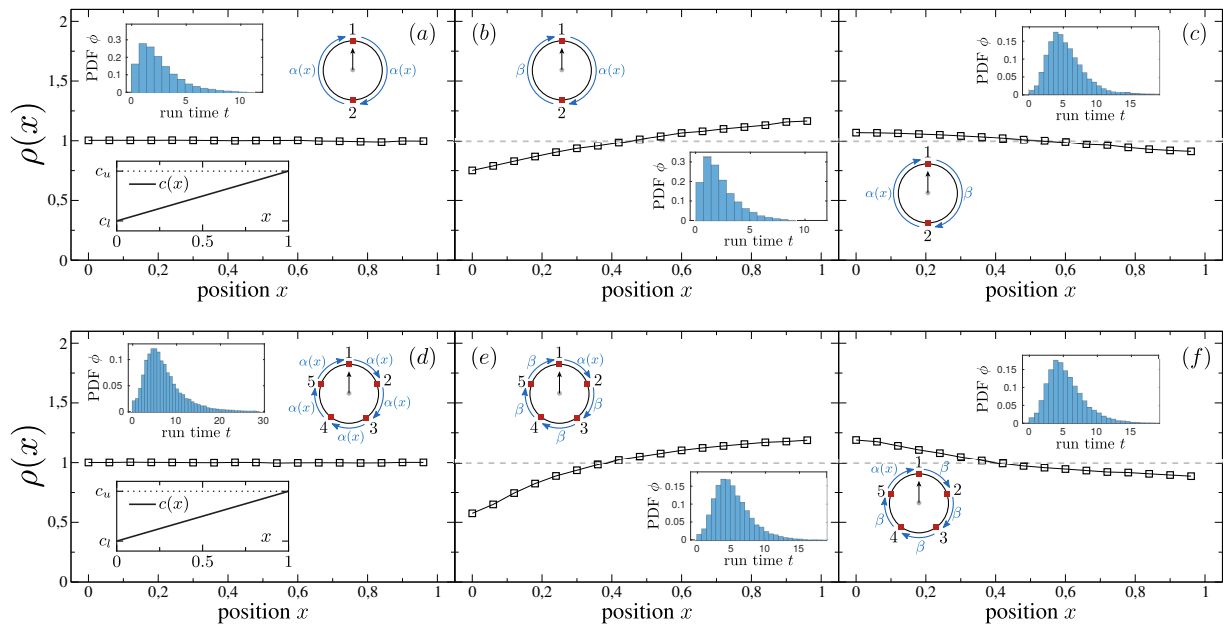

FIG. B.1. Numerical simulations of particles in a one-dimensional box with reflecting boundary conditions using a linear gradient as done in Fig. 3 in the main text, but with a tumbling mechanism instead of reversals: the probability to reverse is $p = 1/2$, and $q = 1/2$ is the probability to keep the same direction after changing from tick $M$ to tick 1. We use $\alpha(x) = c(x)$, $c(x) = x/L$, $c_l = 0$, $c_u = 1$, $\beta = 1$, $v_0 = 0.01$ and $L = 1$.

### 1. Clock-controlled tumbling in one dimension

We implement tumbling in one dimension as follows: When a transition between tick $M$ to 1 occurs, we flip the direction of motion of the particle with a probability of $p = 1/2$ and we keep it as it is with the probability $q = 1 - p$. Fig. B.1 displays the stationary profiles obtained in simulations with this one-dimensional tumbling mechanism. The similarity with the reversal dynamics is evident (cf. Fig. 3 of the main text). The results show that nontrivial stationary profiles appear independently of the mechanism how the direction of motion changes (reversal vs. tumbling), thereby providing a proof of concept that clock-controlled tumbling allows for guided transport and chemotaxis.

### 2. Clock-controlled tumbling in two dimensions

The appearance of the profile $\rho(x)$ as a result of the symmetry breaking of the internal clock of particles is not particular for the one-dimensional model studied in the main text. To illustrate this fact, we implemented a two-dimensional version of our model using individual-based simulations.

A nontrivial profile also emerges in two dimensions whenever an inhomogeneous clock is used. In two-dimensions, the motility is modeled by the differential equation

$$\dot{\mathbf{x}}(t) = v_0 \, \widehat{\mathbf{e}}(t), \tag{B1}$$

where $\mathbf{x}(t)$ is the position of the particle, $\widehat{\mathbf{e}}(t) = (\cos\varphi(t), \sin\varphi(t))$ is a unit vector pointing in the direction of motion and $v_0$ is a constant. The direction of motion $\varphi(t)$ is changed each time the transition from clock tick $M$ to 1 occurs. A new direction of motion $\varphi$ is picked up at random from the uniform probability distribution $p(\varphi) = 1/(2\pi)$ on the interval $[0, 2\pi]$.

The obtained results are shown in Fig. B.2 for the transition rate $\alpha(x, y) = c(x, y) = x/L$, and $\beta = 1$. As observed in the one-dimensional version of the model, the emergence of a taxis response also requires the use of an inhomogeneous clock in two dimensions.

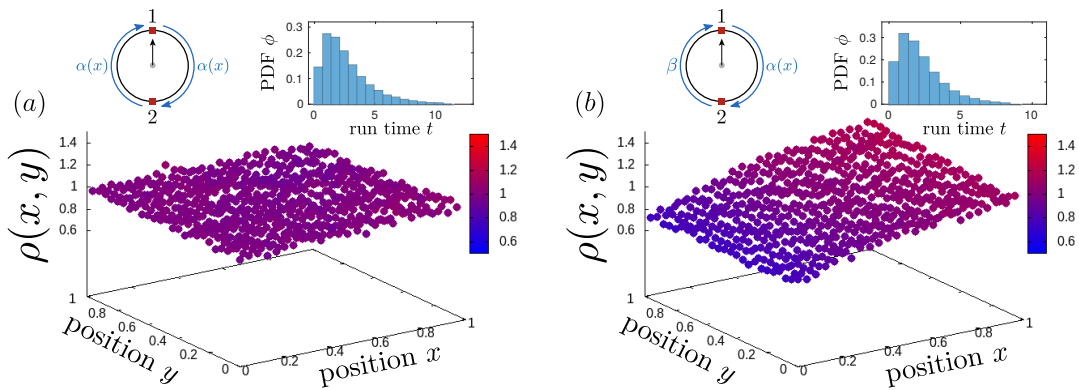

FIG. B.2. Stationary probability densities $\rho(x,y)$ obtained from individual-based simulations using $N = 10^4$ particles that move at constant speed $v_0 = 0.01$ inside a box of dimensions $L \times L$ (with $L = 1$) with reflecting boundary conditions. The left panel displays the stationary profile $\rho(x,y)$ that emerges for a $M = 2$ clock with $\gamma_1(x,y) = \gamma_2(x,y) = \alpha(x,y)$. The right panel corresponds to an inhomogeneous $M = 2$ clock with $\gamma_1(x,y) = \alpha(x,y)$ and $\gamma_2(x,y) = \beta$, with $\beta = 1$. In both panels, $\alpha(x,y) = c(x,y) = x/L$. For each panel, the run-time distribution is plotted as an inset.

## Appendix C: Perturbation theory

### 1. Weakly modulated transition rates

In this section, we explain the mathematical details of the perturbation theory which can be used to approximate the stationary solution of the Master equation corresponding to the model under consideration. Analogous to the notation in the main text, we abbreviate the probabilistic dynamics as follows:

$$\partial_t \mathbf{P}(x,t) = \mathcal{C} \cdot \partial_x \mathbf{P}(x,t) + \mathcal{L}(x) \cdot \mathbf{P}(x,t). \tag{C1}$$

The components of the vector

$$\mathbf{P}(x,t) = \left(P_1^+(x,t), P_1^-(x,t), P_2^+(x,t), P_2^-(x,t), \ldots, P_M^+(x,t), P_M^-(x,t)\right)^T \tag{C2}$$

are the probability density functions to find a particle in position $x$ at time $t$ with an internal state $n$, where $n \in \{1, 2, \ldots, M\}$, moving from left to right $(+)$ or, conversely, from right to left $(-)$. The matrix $\mathcal{C}$ encodes the motility of particles and $\mathcal{L}(x)$ denotes space-dependent transition rates constituting the clock dynamics. The former matrix $\mathcal{C}$ is diagonal and its components read $\mathcal{C}_{ij} = (-1)^j v_0 \delta_{ij}$. The concrete form of $\mathcal{L}(x)$ depends on the specific clock model under consideration. As an example, we specifically state these matrices for the clock model shown in Fig. 3b in the main text:

$$\mathcal{C} = \begin{pmatrix} -v_0 & 0 & 0 & 0 \\ 0 & v_0 & 0 & 0 \\ 0 & 0 & -v_0 & 0 \\ 0 & 0 & 0 & v_0 \end{pmatrix}, \quad \mathcal{L}(x) = \begin{pmatrix} -\alpha(x) & 0 & 0 & \beta \\ 0 & -\alpha(x) & \beta & 0 \\ \alpha(x) & 0 & -\beta & 0 \\ 0 & \alpha(x) & 0 & -\beta \end{pmatrix}. \tag{C3}$$

The dynamics has to be supplemented by appropriate boundary conditions as well as the corresponding initial distribution of particles. Since we will derive the stationary solution only, the initial condition is of minor importance in this context. In this section, we will study periodic boundary conditions only, i.e., we consider all space-dependent functions to be periodic, where the linear system size $L$ determines the period. We clarify in Appendix C 2 that this approach can be directly applied to reflecting boundary conditions as well.

The central assumption of the perturbation theory is that the position dependence of the transition rates is weak. More precisely, we split the matrix $\mathcal{L}(x)$ into a constant and a spatially dependent part according to

$$\mathcal{L}(x) = \mathcal{L}_0 + \varepsilon \Delta(x) \mathcal{L}_S, \tag{C4}$$

where the scalar function $\Delta(x)$ encodes the spatial dependence of the rates, the constant matrix $\mathcal{L}_S$ states which of the components are spatially modulated and $\varepsilon$ is defined by the condition:

$$0 = \int_{-L/2}^{+L/2} dx\, \Delta(x). \tag{C5}$$

For the specific clock model mentioned above, the components of the matrices $\mathcal{L}_0$ and $\mathcal{L}_S$ read

$$\mathcal{L}_0 = \begin{pmatrix} -\alpha_0 & 0 & 0 & \beta \\ 0 & -\alpha_0 & \beta & 0 \\ \alpha_0 & 0 & -\beta & 0 \\ 0 & \alpha_0 & 0 & -\beta \end{pmatrix}, \quad \mathcal{L}_S = \begin{pmatrix} -1 & 0 & 0 & 0 \\ 0 & -1 & 0 & 0 \\ 1 & 0 & 0 & 0 \\ 0 & 1 & 0 & 0 \end{pmatrix}. \tag{C6}$$

Henceforth, we consider $\varepsilon$ to be small such that the solution of the Master equation can be expanded in a Taylor series in this expansion parameter. In short, we propose the ansatz

$$\mathbf{P}(x,t) = \sum_{\mu=0}^{\infty} \mathbf{P}_\mu(x,t) \varepsilon^\mu. \tag{C7}$$

Inserting into Eq. (C1) yields the following conditional equations for $\mathbf{P}_\mu(x,t)$:

$$\partial_t \mathbf{P}_\mu(x,t) = \mathcal{C} \cdot \partial_x \mathbf{P}_\mu(x,t) + \mathcal{L}_0 \cdot \mathbf{P}_\mu(x,t) + \Delta(x) \cdot \begin{cases} 0, & \mu = 0, \\ \mathcal{L}_S \cdot \mathbf{P}_{\mu-1}(x,t), & \mu \geq 1. \end{cases} \tag{C8}$$

In the following, we sketch the derivation of the solution of these equations to the lowest, nontrivial order.

In the lowest order perturbation theory, all transition rates are constant and, hence, the stationary state $\mathbf{P}_0$ is spatially homogeneous. It is thus easily found from the condition $\mathcal{L}_0 \cdot \mathbf{P}_0 = 0$ which allows us to determine $\mathbf{P}_0$ up to a normalization constant that is given by the particle number conservation – the sum over all components of $\mathbf{P}_0$ must equal the uniform probability distribution function $\rho_0 = 1/L$.

The stationary solution of the first, nontrivial order is determined by the following differential equation:

$$\left( -\mathcal{C} \frac{d}{dx} - \mathcal{L}_0 \right) \cdot \mathbf{P}_1(x) = \Delta(x) \mathcal{L}_S \cdot \mathbf{P}_0. \tag{C9}$$

There is a mathematical subtlety since this equation does not possess a unique solution as the operator on the left hand side is not invertible. To see this, let $\mathbf{P}_1(x) = \bar{\mathbf{P}}_1(x)$ be particular solution. Consequently, the function $\mathbf{P}_1(x) = \bar{\mathbf{P}}_1(x) + c \mathbf{P}_0$ is a solution as well for all real coefficients $c \in \Re$. This ambiguity is resolved by reconsidering the normalization condition as detailed below. First, we integrate Eq. (C9) over the entire system yielding

$$\frac{1}{L} \int_{-L/2}^{+L/2} dx \left( -\mathcal{C} \frac{d}{dx} - \mathcal{L}_0 \right) \cdot \mathbf{P}_1(x) = -\mathcal{L}_0 \cdot \mathbf{p}_1 = 0, \tag{C10}$$

where $\mathbf{p}_\mu = L^{-1} \int_{-L/2}^{+L/2} dx \, \mathbf{P}_\mu(x)$ was introduced. Therefore, we obtain the intermediate result $\mathbf{p}_1 \propto \mathbf{p}_0$, since $\mathcal{L}_0 \cdot \mathbf{P}_0 = 0$ and, thus, $\mathcal{L}_0 \cdot \mathbf{p}_0 = 0$. However, we have already normalized the lowest order solution $\mathbf{p}_0$ correctly – the sum over all components of $\mathbf{p}_0$ equals $1/L$ – and, consequently, we conclude that $\mathbf{p}_1$ must equal zero. In short, the first order correction $\mathbf{P}_1(x)$ has to be determined such that the integral over this function vanishes,

$$0 = \int_{-L/2}^{+L/2} dx \, \mathbf{P}_1(x), \tag{C11}$$

thereby ensuring the solvability of Eq. (C9) unambiguously. The explicit solution

$$\mathbf{P}_1(x) = \frac{1}{L} \left[ \sum_{\substack{n=-\infty, \\ n \neq 0}}^{\infty} \Delta_n e^{-iq_n x} \mathcal{G}(q_n) \right] \cdot \mathcal{L}_S \cdot \mathbf{P}_0 \tag{C12a}$$

$$= \frac{2}{L} \left\{ \sum_{n=1}^{\infty} \Re \Big[ \mathcal{G}(q_n) \Delta_n \Big] \cos(q_n x) + \Im \Big[ \mathcal{G}(q_n) \Delta_n \Big] \sin(q_n x) \right\} \cdot \mathcal{L}_S \cdot \mathbf{P}_0 \tag{C12b}$$

is found by Fourier transformation of Eq. (C9) with respect to the spatial variable $x$. It is easily verified by inserting into this equation. In the expression above, we abbreviated the wavenumbers $q_n = 2\pi n/L$ with $n \in \mathbb{Z} \backslash \{0\}$, the Fourier coefficients $\Delta_n$ of the spatial modulation $\Delta(x)$ defined by

$$\Delta_n = \int_{-L/2}^{+L/2} dx \, e^{iq_n x} \Delta(x) \tag{C13}$$



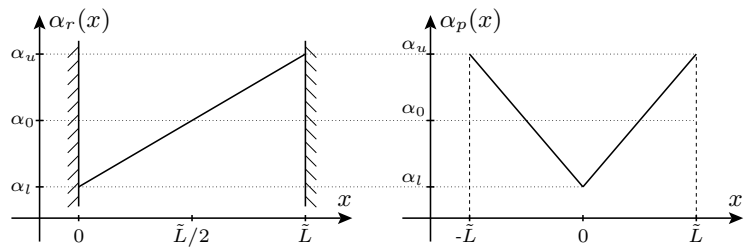

FIG. C.1. Illustration of the analogy of reflecting boundary conditions (left) and a corresponding system with periodic boundary conditions which has the double length with respect to the former. In the stationary state, the probability distribution $\mathbf{P}(x)$ on the interval $x \in (0, \tilde{L})$ will be similar if $\alpha_p(x)$ is constructed from $\alpha_r(x)$ by mirror reflection, cf. Eq. (C15).

as well as the propagator $\mathcal{G}(q) = (iq\mathcal{C} - \mathcal{L}_0)^{-1}$, defined for $q \neq 0$.

In short, the condition $\mathcal{L}_0 \cdot \mathbf{P}_0 = 0$ in combination with the expression for the first order, nontrivial correction $\mathbf{P}_1(x,t)$, cf. Eq. (C12), determine the stationary solution to first order

$$\mathbf{P}(x) = \mathbf{P}_0 + \varepsilon \mathbf{P}_1(x) + \mathcal{O}(\varepsilon^2), \tag{C14}$$

which can be rewritten for a specific clock model by specification of $\Delta(x)$ as well as the matrices $\mathcal{L}_0$ as well as $\mathcal{L}_S$ as explained in Appendix C 3.

### 2. Reflecting vs. periodic boundary conditions

So far, we considered a perturbation theory for stationary density profiles which is applicable to systems with periodic boundary conditions. However, one can utilize the expressions derived above also for systems with reflecting boundary conditions as discussed in the following, cf. Fig. C.1. For the sake of clarity, we consider a linear system [system size $\tilde{L}$ and $x \in (0, \tilde{L})$] where one of the transition rates, denoted by $\alpha_r(x)$, depends explicitly on space. Now the question is: what happens at the boundaries of the system? If a particle, moving from right to left, touches the boundary in $x = 0$, it will be reflected. In other words, a particle, which was moving into the direction of decreasing transition rates $\alpha_r(x)$, will move towards increasing transition rates after the reflection. Now, this situation may analogously be observed in a periodic system which is as twice as long [$x \in (-\tilde{L}, \tilde{L})$], where the space dependent transition rate is replaced by

$$\alpha_p(x) = \begin{cases} \alpha_r(+x), & x \geq 0, \\ \alpha_r(-x), & x < 0. \end{cases} \tag{C15}$$

In the stationary state, the number of particles passing through the point $x = 0$ from left to right is equal to the number of particles passing this point moving in the opposite direction due to the symmetry of the problem. Therefore, the particle flux in $x = 0$ vanishes. This line of arguments holds similarly for the other boundary at $x = \tilde{L}$. Accordingly, the dynamics of the two systems, one with reflecting and the extended system with periodic boundary conditions, is completely equivalent on the interval $x \in (0, \tilde{L})$. In short, we can map every system with reflecting boundary conditions onto an extended system with periodic boundary conditions and a space dependent transition rate $\alpha_p(x)$ that is constructed by mirroring $\alpha_r(x)$ as expressed by Eq. (C15).

### 3. An example

To fix ideas, we study the particular solution for the clock model shown in Fig. 3b in the main text for a linear gradient as it is represented in Fig. C.1. The matrices $\mathcal{L}_0$ and $\mathcal{L}_S$ were already given explicitly in Appendix C 1. The lowest order solution, to be determined from $\mathcal{L}_0 \cdot \mathbf{P}_0$, reads

$$\mathbf{P}_0 = \frac{1}{2\tilde{L}(\alpha_0 + \beta)} \begin{pmatrix} \beta \\ \beta \\ \alpha_0 \\ \alpha_0 \end{pmatrix}. \tag{C16}$$

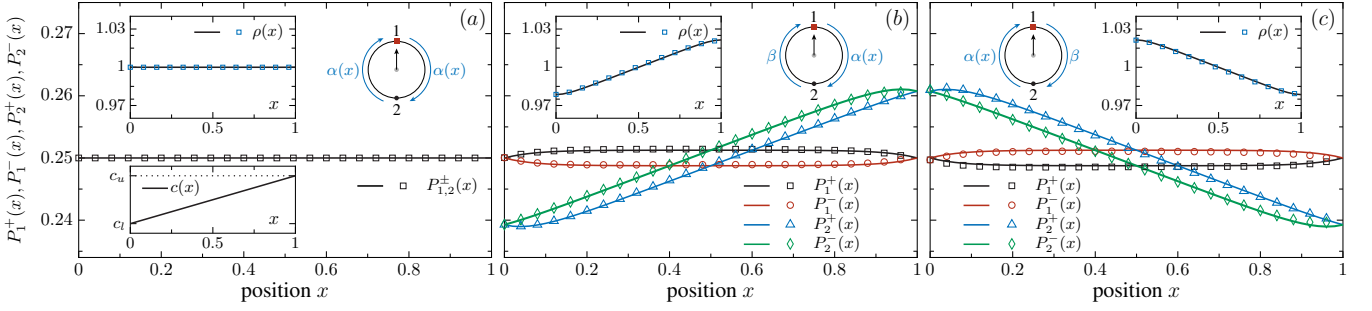

FIG. C.2. Stationary distributions for the same motifs presented in Fig. 3 of the main text with M = 2. Points represent particle-based simulations and lines show the density profiles as predicted by the perturbation theory. In the large panels, the stationary probability density $P_n^\pm(x)$ to find a particle at position x with orientation $s = \pm 1$ in the internal state $n$ is plotted. Insets represent the stationary probability density $\rho(x)$ obtained from the sum over all components $P_n^\pm(x)$. The external field $c(x)$, common for (a), (b) and (c), is shown as an inset in panel $(a)$. Parameters in arbitrary units: $\alpha(x) = \alpha_b(1 + c(x))$, $\alpha_b = 9.5$, $c(x) = 0.1x/L$, $c_l = 0$, $c_u = 0.1$, $\beta = 10$, $v_0 = 1$, $M = 2$, $L = 1$. For particle-based simulations, $N = 10^4$ particles were simultaneously tracked using a stochastic Euler scheme with $\Delta t = 10^{-3}$ and density distributions were averaged over time neglecting initial transients.

The parameter $\alpha_0$ denotes the mean transition rate $\alpha_0 = (\alpha_l + \alpha_r)/2$, i.e., $\alpha_r(x) = \alpha_0 + \varepsilon\Delta(x)$. Further, we parametrize the spatial modulation $\Delta(x)$ according to

$$\Delta(x) = -1 + \frac{2x}{\tilde{L}}. \tag{C17}$$

Now, we can make immediate use of the equations derived in Appendix C 1, however, care must be taken that the system size was reparametrized: $\tilde{L} = L/2$. Accordingly, the wavenumbers $q_n$ are determined by $q_n = 2\pi n/L = \pi n/\tilde{L}$. To utilize Eq. (C12), we need to calculate the Fourier coefficients $\Delta_n$ corresponding to the spatial modulation $\Delta(x)$, which read

$$\Delta_n = \int_{-L/2}^{+L/2} dx\, e^{iq_n x} \Delta(x) = -\frac{2L}{n^2\pi^2}\left[1 - (-1)^n\right] \tag{C18}$$

for $n \neq 0$ and $\Delta_0 = 0$, consistently with the condition (C5). Due to the symmetry $\Delta(-x) = \Delta(x)$ of the spatial dependence of the transition rate, these Fourier coefficients are real numbers. What is left to be done is the calculation of the propagator $\mathcal{G}(q) = (iq\mathcal{C} - \mathcal{L}_0)^{-1}$ – obtained from the inversion of a $(4 \times 4)$-matrix – and insertion of the result into Eq. (C12), which yields after some straightforward matrix algebra

$$\mathbf{P}(x) = \frac{1}{L(\alpha_0 + \beta)} \cdot \left\{ \begin{pmatrix} \beta \\ \beta \\ \alpha_0 \\ \alpha_0 \end{pmatrix} + \varepsilon \cdot \frac{2}{L} \sum_{n=1}^{\infty} \frac{\beta \Delta_n}{\alpha_0^2 + \beta^2 + (q_n v_0)^2} \left[\cos(q_n x)\begin{pmatrix} \beta - \alpha_0 \\ \beta - \alpha_0 \\ \beta + \alpha_0 \\ \beta + \alpha_0 \end{pmatrix} + q_n v_0 \sin(q_n x)\begin{pmatrix} -1 \\ +1 \\ +1 \\ -1 \end{pmatrix}\right] \right\} + \mathcal{O}(\varepsilon^2). \tag{C19}$$

Alternatively, one may rewrite the stationary probability density to first order in perturbation theory by insertion of the Fourier modes $\Delta_n$ as well as the original system size $\tilde{L}$:

$$\mathbf{P}(x) \simeq \frac{1}{2\tilde{L}(\alpha_0 + \beta)} \cdot \left\{ \begin{pmatrix} \beta \\ \beta \\ \alpha_0 \\ \alpha_0 \end{pmatrix} - \varepsilon \cdot \frac{8}{\pi^2} \sum_{n=0}^{\infty} \frac{\beta/(2n+1)^2}{\alpha_0^2 + \beta^2 + (q_{2n+1}v_0)^2} \left[\cos(q_{2n+1}x)\begin{pmatrix} \beta - \alpha_0 \\ \beta - \alpha_0 \\ \beta + \alpha_0 \\ \beta + \alpha_0 \end{pmatrix} + q_{2n+1}v_0 \sin(q_{2n+1}x)\begin{pmatrix} -1 \\ +1 \\ +1 \\ -1 \end{pmatrix}\right] \right\}. \tag{C20}$$

The stationary probability distribution for the position of a particle, irrespective of its internal state, is eventually given by the sum over all components of the vector $\mathbf{P}(x)$. The perturbation theory results are compared to individual

6

based simulations in the main text.

---



\end{document}